\pdfoutput=1

\documentclass[3p]{elsarticle}

\newcommand{\diff}{\ \mathrm{d}}
\newcommand{\sdiff}{\mathrm{d}}

\newcommand*{\thead}[1]{\multicolumn{1}{c}{#1}}

\usepackage{savesym}
\savesymbol{tablenum}
\usepackage[exponent-product=\cdot]{siunitx}

\usepackage{subcaption}
\usepackage{amsmath}
\usepackage{calc}
\usepackage{aas_macros}
\begin{document}


\title{AGN Neutrino flux estimates for a realistic hybrid model}


\author[nwu]{S. Richter}
\ead{Stephan.Richter@nwu.ac.za}

\author[nwu]{F. Spanier}
\ead{felix@fspanier.de}
\address[nwu]{Centre for Space Research\\
North-West University\\
2520 Potchefstroom, South Africa}


\begin{abstract}
Recent reports of possible correlations between high energy neutrinos observed by IceCube and Active Galactic Nuclei (AGN) activity sparked a burst of publications that attempt to predict the neutrino flux of these sources. However, often rather crude estimates are used to derive the neutrino rate from the observed photon spectra. In this work neutrino fluxes were computed in a wide parameter space. The starting point of the model was a representation of the full spectral energy density (SED) of \textit{3C 279}. The time-dependent hybrid model that was used for this study takes into account the full $p\gamma$ reaction chain as well as proton synchrotron, electron-positron-pair cascades and the full SSC scheme. We compare our results to estimates frequently used in the literature. This allows to identify regions in the parameter space for which such estimates are still valid and those in which they can produce significant errors. Furthermore, if estimates for the Doppler factor, magnetic field, proton and electron densities of a source exist, the expected IceCube detection rate is readily available.
\end{abstract}

\begin{keyword}
acceleration of particles -- galaxies: jets -- neutrinos -- radiation mechanisms: non-thermal -- relativistic processes
\end{keyword}

\maketitle

\section{Introduction} \label{sec:introduction}
Since the IceCube detector \citep{2017JInst..12P3012A} provided evidence for a high-energy, extragalactic neutrino flux \citep{2014PhRvL.113j1101A}, it has been one of the large goals of high-energy astrophysics to connect the observed flux to known sources. Although photohadronic models for Blazars have evolved for a long time \citep{1993A&A...269...67M}, the small number of observed high-energy neutrinos makes the detection of a source or even a source population challenging. In addition to stacking analyses \citep{2017ApJ...835...45A}, time-dependent searches were carried out \citep{2015ApJ...805L...5A,2015ApJ...807...46A}, both of which have been unsuccessful so far. In either case, in order to derive a discovery potential from the photon spectrum or interpret a possible non-detection, a robust model is required. However, often extremely simplified models for the relation of the photon and neutrino flux are employed \citep[e.g.][]{2016arXiv160202012K,2016ApJ...831...12H}. Without a proper understanding of the validity of the employed assumptions, an interpretation of such models is rather difficult.

Furthermore, with the advances in computational power, much more complex models can produce results on single nodes within times of the order of minutes. Although we admit the difficulties in the analysis of large samples of sources, especially bright and strongly variable sources should be analyzed more rigorously. The work at hand aims to show that such an analysis is indeed easily possible and highly necessary. For this purpose we compare fluxes and neutrino rates, computed consistently within our model, with predictions of models recently used in the literature using the modeled spectral energy density (SED) as input.

The model used for this work is fully time-dependent and reflects the complex interplay of electron and proton synchrotron emission, photo meson production as well as the subsequent decays and pair-cascades. It is described in section~\ref{sec:model}.

In section~\ref{sec:results} first an approximate fit to the SED of \textit{3C 279} is presented, which serves as a starting point for the parameter scan, whose results are presented subsequently. Even simple hadronic AGN models make use of a large number of parameters. It is computationally not possible to scan the entire high-dimensional parameter-space. Therefore we focus on three parameters that presumably are the most important ones. Section ~\ref{sec:discussion} will discuss the obtained results and discrepancies between the analyzed models. A summary of our conclusions and a brief outlook on how complex modeling can become more mainstream are presented in section~\ref{sec:conclusion}.

\section{Model}
\label{sec:model}
The characteristic double hump structure found in the SEDs of Blazars can often be explained elegantly by the SSC paradigm. However, it was shown that the SEDs of low peaked BL Lacs and FSRQs can often also be explained by a combination of electron synchrotron emission, proton synchrotron emission and the photo-hadronic reaction chain, provided higher magnetic fields that confine the high-energy protons~\citep[e.g.][]{2013EPJWC..6105009W}. Those so called hybrid-models allow an unbiased modeling of many sources, from SSC dominated to strongly hadronic. They are also ideal tools for studies identifying parameter regions with a high neutrino efficiency, i.e. a large ratio of produced neutrinos over emitted photons.

Historically the model at hand emerges from a combination of the models described in \cite{2013EPJWC..6105009W} and \cite{2016ApJ...829...56R}. During this process the two independently developed code bases were also reviewed against each other. For an efficient computation we choose a homogeneous description adapting the two-zone geometry introduced by \cite{2013EPJWC..6105009W}, consisting of a spherical radiation zone with a radius $R_{\text{rad}}$ and a smaller, nested acceleration zone with radius $R_{\text{acc}}$\footnote{It should be noted that in time-independent analyses, this approach is equivalent to a homogeneous one zone model into which a fully developed power-law is injected.}. Monoenergetic protons and electrons are injected into the acceleration zone and  gain energy until the synchrotron losses balance any further acceleration gains. Particles eventually escape into the radiation zone, where they undergo all implemented radiation and scattering processes, introducing further species, namely pions, muons, positrons and neutrinos. Except for the extremely short lived pions, all species are treated as time-dependent particle distributions.

The routines computing the leptonic radiation processes are taken from \cite{2016ApJ...829...56R}, dropping all spatial dependencies. A time-dependent implementation of the photo-hadronic processes follows the model \textit{Sim-B} of \cite{2010ApJ...721..630H}. All internal sources of radiation contribute to the target photon field, that is electron-, proton- and muon synchrotron, $\pi_0$-decay and the synchrotron pair-cascades. External photon fields are not considered.

The particle acceleration used here is closely related to the statistical approach of \cite{2016ApJ...829...56R}. Due to the omitted spatial dependency, a simplified approach was developed, closely following \cite{2004PASA...21....1P}. Diffusion of particles across a shock front is assumed to happen within the acceleration zone, leading to Fermi-I acceleration. The diffusion parameter $\eta$ is defined per particle species and is assumed to scale linearly with the mass. This allows to compute the escape timescale for the acceleration zone
\begin{equation}
    T_{\text{esc}}=\frac{3}{4}\frac{\eta R_{\text{acc}}}{c}\quad.
\end{equation}
The probabilities for a particle to either return to the shock ($P_{\text{acc}}$) or escape further downstream and hence into the radiation zone ($P_{\text{esc}}$) are computed as~\citep{2004PASA...21....1P}
\begin{align}
 P_{\text{esc}}&=\frac{4V_\mathrm S}{r}\quad,\\
 P_{\text{acc}}&=1-P_{\text{esc}}\quad,
\end{align}
with the shock speed $V_\mathrm S$ and the compression ratio $r$. The energy gain by shock crossing therefore happens on a time scale
\begin{equation}
 T_{\text{acc}}=T_{\text{esc}}\frac{P_{\text{esc}}}{P_{\text{acc}}}
\end{equation}
with an average energy gain of ~\citep{2004PASA...21....1P}
\begin{equation}
 \frac{\Delta E}{E}=1+\frac{4}{3}\frac{r-1}{r}V_\mathrm S\quad.
\end{equation}
The acceleration is then implemented as explicit scatterings between energy bins. In comparison to an analytic acceleration term as part of the Fokker-Planck equation, this ansatz avoids the artificial cut-off at a set energy $\gamma_{\text{max}}$ (and the accompanying numerical difficulties) and correctly reflects the exponential decay of the particle distribution at $\gamma>\gamma_{\text{max}}$ due to the stochastic nature of the assumed acceleration process.

Eventually the photon and neutrino fluxes can be computed from the photon density $N_{\tilde\nu}(\tilde\nu)$ and neutrino density $M_{\tilde E}(\tilde E)$, respectively\footnote{In order to clearly distinguish the differential densities and fluxes of photons and neutrinos, we here use the uncommon variables $M$ and $G$ for neutrinos. Please note that for absolute numbers used in section \ref{sec:results}, we keep the variables $N_\gamma$ and $N_\nu$, respectively.}, employing
\begin{equation}
  \nu F_\nu(\nu)=\frac{\mathcal{D}^4}{1+\mathcal Z}\frac{h\tilde\nu^2c}{4}\frac{R_{\text{rad}}^2}{d_l^2}N_{\tilde\nu}(\tilde\nu)\quad\quad E\cdot G_E(E)=\frac{\mathcal{D}^4}{1+\mathcal Z}\frac{\tilde E^2c}{4}\frac{R_{\text{rad}}^2}{d_l^2}M_{\tilde E}(\tilde E)\quad,
\end{equation}
with the Doppler factor $\mathcal{D}$, redshift $\mathcal Z$ and luminosity distance $d_l$. The neutrino density is boosted the same way. The frequency and energy, respectively, at which the flux is observed has to be boosted according to $\nu=\mathcal{D}\tilde\nu/(1+\mathcal Z)$ and $E=\mathcal{D}\tilde E/(1+\mathcal Z)$.

In summary the model requires 10 input parameters, summarized in Table~\ref{tab:parameters}. For the sake of simplicity the spatial diffusion parameter is fixed to $\eta=1$. It should be noted that as it is common to any lepto-hadronic model \citep{1993A&A...269...67M,2001APh....15..121M} the electron and proton densities describe solely the high-energy population and charge neutrality is given when low energy populations are accounted for.

\begin{table}[h]
	\centering
	\caption{Summary of parameters.}
	\begin{tabular}{clcc}
		\hline\hline
		\thead{parameter} & \thead{description} & \thead{fit} & \thead{range}\\
		\hline
		\\[-3mm]
		$R_{\text{acc}}$ & radius of acceleration zone & $\SI{1E14}{cm}$ & -\\
		$R_{\text{rad}}$ & radius of radiation zone & $\SI{1E16}{cm}$ & -\\
		
		$B$ & magnetic field strength & $\SI{5}{G}$ & $\{1,5,25\}\,\SI{}{G}$\\
		
		$\gamma_{\text{inj,p}}$ & injection energy for protons & $\SI{100}{}$ & -\\
		$\gamma_{\text{inj,el}}$ & injection energy for electrons & $\SI{100}{}$ & -\\
		$N_{\text{inj,p}}$ & injection rate for protons & $\SI{1E46}{s^{-1}}$ & $\SI{2e42}{}-\SI{2e46}{s^{-1}}$ \\
		$N_{\text{inj,el}}$ & injection rate for electrons & $\SI{5E44}{s^{-1}}$ & $\SI{1e42}{}-\SI{1e46}{s^{-1}}$\\
		
		$V_\mathrm S$ & shock speed & $\SI{0.1}{c}$ & -\\
		$r$ & shock compression ratio & $3$ & -\\
		
		$\mathcal{D}$ & Doppler factor & $40$ & -\\
		\hline
	\end{tabular}
	\label{tab:parameters}
\end{table}

\section{Results}
\label{sec:results}
\begin{figure}[ht]
  \centering
  \includegraphics[width=\textwidth]{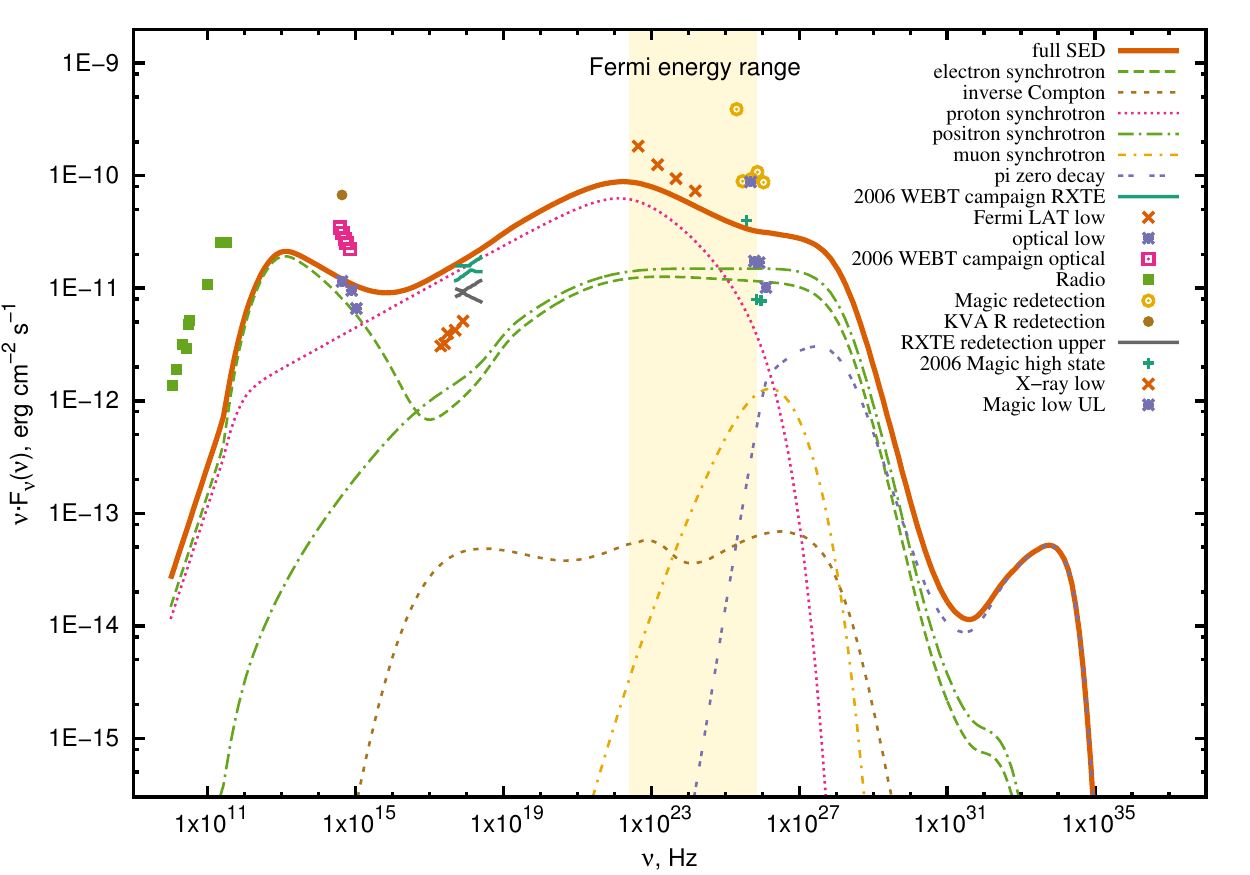}
  \caption{An approximate fit of the hybrid model to the SED of \textit{3C 279}. Data taken from~\citep{2011A&A...530A...4A}. We use the de-absorbed data as given in the paper.}
  \label{fig:base_sed}
\end{figure}

The starting point for the parameter scan is an approximate fit to the SED of \textit{3C 279}. The SED is presented in Fig~\ref{fig:base_sed} and the fit parameters are shown in Table~\ref{tab:parameters}. The fit was done ``by eye'' and only aims to qualitatively model the distinct features of the SED, namely the flux scaling, spectral indices and energy cutoffs.
The integral over the computed neutrino fluxes yields an IceCube detection rate\footnote{Neutrino absorption by matter can be ignored completely in this scenario the approximate gramage is on the order of 1 g/cm$^2$ yielding not neutrino interaction} of $dN_{IC}/dt=\SI{3.08E-8}{s^{-1}}$ or $\SI{2.6}{}$ events above $\SI{60}{TeV}$ in $988$ days. Considering that even the base SED data used here represents a state with unusual high flux, this is in line with the fitted number of events $n_s=1{.}1$ from~\citep{2016ApJ...823...65A}. The expected background of events with $E_{dep}>\SI{60}{TeV}$ from the declination of \textit{3C 279} is around one event in $988$ days~\citep{2014PhRvL.113j1101A} \footnote{The largest observed flare of the source~\citep{2015ApJ...808L..48P} reaches a flux of several $\SI{E-9}{erg\,cm^{-2} s^{-1}}$ in the Fermi range. This can be reproduced in our model by a fit with $B=\SI{25}{G}$ and a proton injection rate of $N_{inj,p}=\SI{2e46}{s^{-1}}$, yielding a neutrino rate of $285$ IceCube events in $988$ days above $\SI{60}{TeV}$. In order to derive a meaningful number, one would have to know the duty cycle of 3C279.}.

Subsequently to obtaining this fit, the electron and proton densities are altered over four orders of magnitude on a grid of size $30\times30$. This is done for three different values of the magnetic field strength $B=\{1,5,25\}\,\mathrm{G}$. Geometric parameters like jet bending or Doppler boosting are not altered in this model as they affect photons and neutrinos in the same way yielding no change their ratio.
\subsection{Verification}
\label{sec:verification}
\begin{figure*}[t!]
    \centering
    \begin{subfigure}[t]{0.31\textwidth}
        \centering
        \includegraphics[height=1.7in]{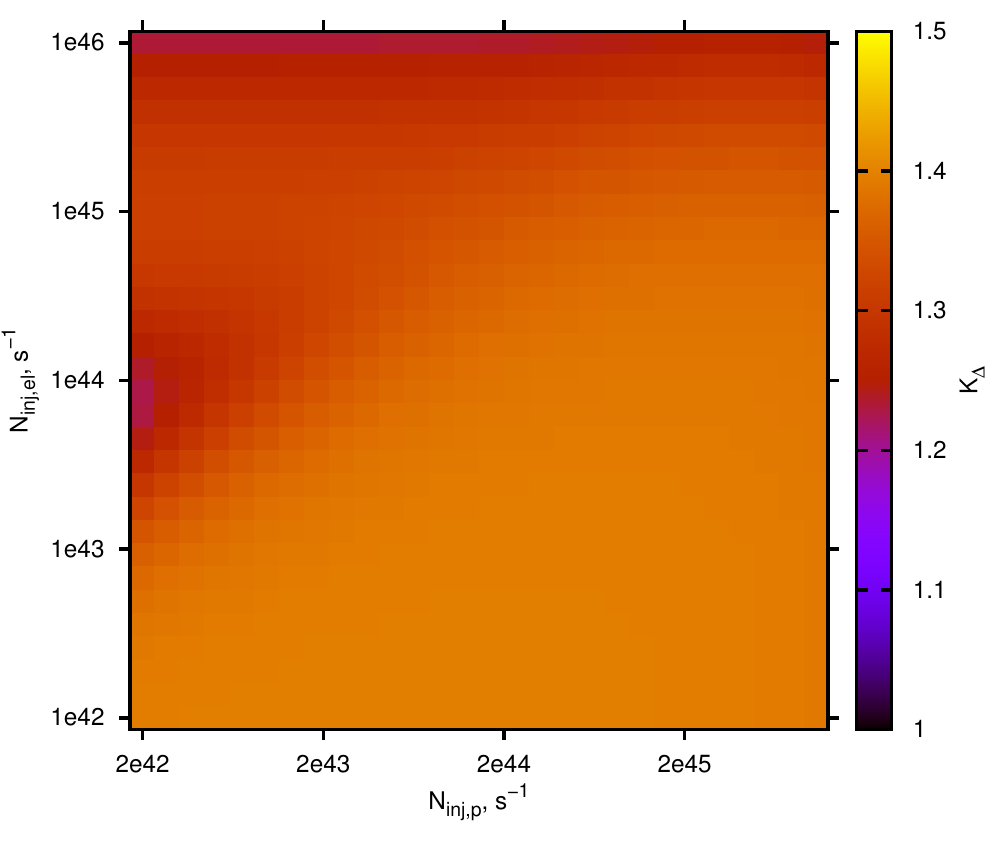}
        \caption{Factor $K$ from Eq. \ref{eq:factor_k} with $N_\gamma$ the number of photons produced in $\pi_0$-decay and $N_\nu$ the number of neutrinos produced via the $\delta$-resonance.}
    \end{subfigure}
     ~
    \begin{subfigure}[t]{0.31\textwidth}
        \centering
        \includegraphics[height=1.7in]{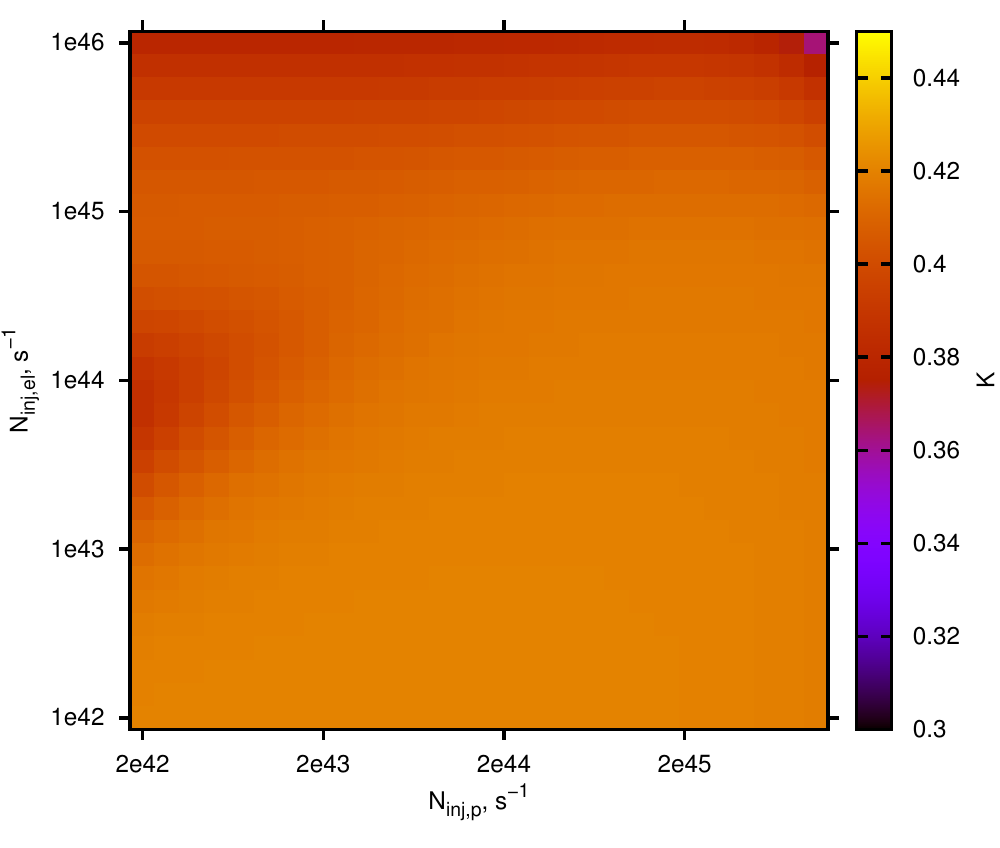}
        \caption{Factor $K$ from Eq. \ref{eq:factor_k} with $N_\gamma$ the number of photons produced in $\pi_0$-decay and $N_\nu$ the number of neutrinos produced in all photohadronic reaction chains.}
    \end{subfigure}%
    ~
    \begin{subfigure}[t]{0.31\textwidth}
        \centering
        \includegraphics[height=1.7in]{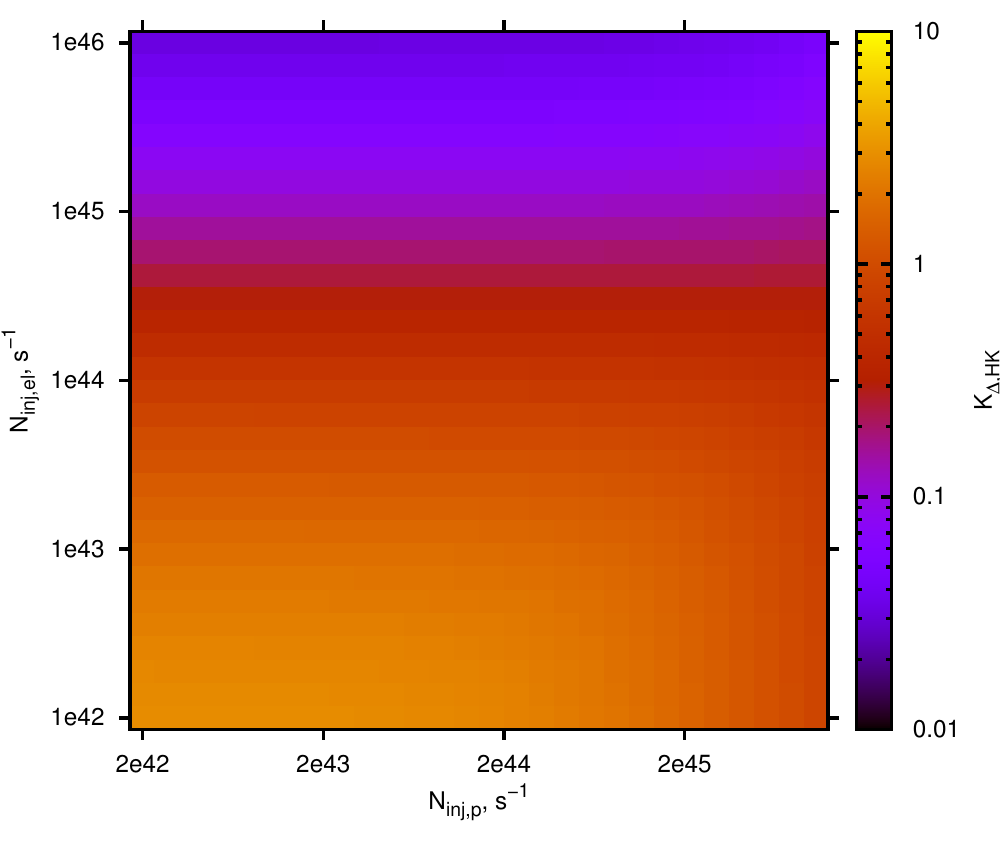}
        \caption{Like panel (a), but using the definition from \citep[Eq. (1)][]{2016ApJ...831...12H}.}
    \end{subfigure}
    \caption{Comparison of different definitions of the factor $K$ at $B=\SI{1}{G}$.}
    \label{fig:verification}
\end{figure*}
Numerical solutions always posses the possibility of conceptual and implementation errors. Therefore thorough testing is required. Although the full photohadronic reaction chain is implemented in the model used for this study, it is possible to investigate individual contributions. This allows to test our implementation and the correct scaling of the computed fluxes. For this purpose we adopt a relation presented in \citep[Eq. (1)][]{2016ApJ...831...12H}, with $K$ being computed from our simulations.
We would also like to point out what we believe is an error in the expression used in~\cite{2016ApJ...831...12H}, namely that the particle densities should be compared, not the fluxes.

Figure~\ref{fig:verification} panels a) and b)  show the value $K$, computed from the photon current density $N_\gamma=F_\gamma/E_\gamma$ and the neutrino current density $N_\nu=F_\nu/E_\nu$
\begin{equation}
 \label{eq:factor_k}
 \int \frac{\sdiff N_\gamma}{\sdiff E_\gamma}\diff E_\gamma=K\cdot\int \frac{\sdiff N_\nu}{\sdiff E_\nu}\diff E_\nu\quad,
\end{equation}
the integrals being over the entire energy domain. For panel c), we use the same definition as~\cite{2016ApJ...831...12H}.

The expected value for $K$ in case of the $\Delta(1232)$-resonance can be computed form the branching ratio into $\pi^+\ (1/3)$ and $\pi^0\ (2/3)$ and the decay chain of $\pi^+$ \citep[e.g.][]{2010ApJ...721..630H}
\begin{equation}
 \pi^+\rightarrow(e^++\nu_e+\bar{\nu}_\mu)+\nu_\mu\quad.
\end{equation}
When only considering the  $\delta$-resonance, the theoretical value for $K=4/3$ should be reproduced by the model in the entire parameter space. In panel a) this can be seen almost perfectly. The next panel shows that even for multi-pion production, there is no dependence of $k$ on the particle densities, as might be expected. The number of neutrinos produced per photon from photo-hadronic interactions more than triples, resulting in a factor $k\approx0.4$. However, this number will in general depend on other parameters. Also note the deviation from the doubling of the neutrino production rate assumed by~\cite{2016ApJ...831...12H}. Moreover, their comparison of the fluxes breaks down for certain changes in the parameters, as can be seen in panel c). Here an increase in electron density leads to a shift of the target photon population. This in turn shifts the energy dependent branching ratios of the decay of secondary particles, namely the $\pi^+$.

\subsection{Parameter Scan}
In this section we investigate the complex connection between the photon flux in the Fermi range, which serves as a baseline for the majority of studies on candidates of neutrino sources, and the neutrino flux observable by the IceCube detector.

\begin{figure}[ht]
  \centering
  \includegraphics[width=\textwidth]{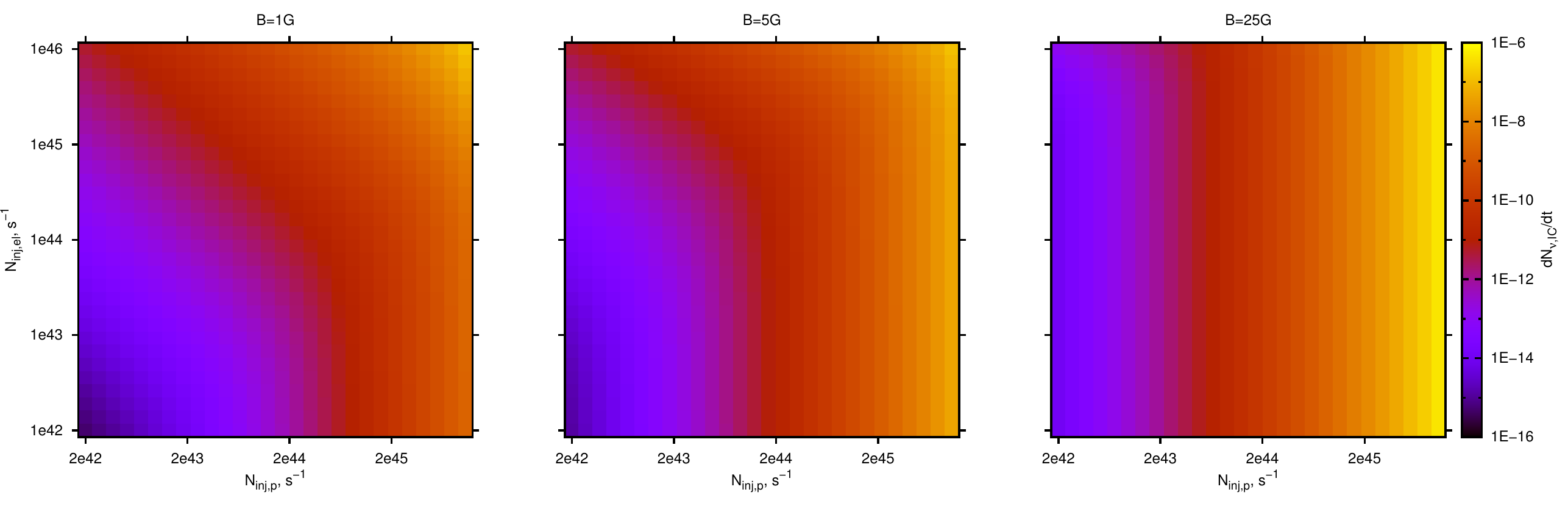}
  \caption{Expected detection rate (counts$\cdot s^{-1}$) from a model source.}
  \label{fig:icecube_detection_rate}
\end{figure}

The probably most interesting number for a neutrino candidate source currently is the expected IceCube detection rate. For our model source this quantity is show in Fig.~\ref{fig:icecube_detection_rate}, for which we only considered muon neutrinos and used the IceCube effective area taken from~\cite{2014ApJ...796..109A}.
For the fit parameters of \textit{3C 279} ($B=\SI{5}{G}$, $N_{\text{inj,p}}=\SI{1E46}{s^{-1}}$, $N_{\text{inj,el}}=\SI{5E44}{s^{-1}}$) a rate of approximately $\SI{e-8}{s^{-1}}$ is computed, equaling one neutrino per three years as might be expected from an extremely bright source like \textit{3C 279}.

Of further interest is the change in the dependency on the particle densities. For low magnetic fields both electron and proton densities almost equally influence the resulting neutrino rate, as would be expected from photo hadronic interactions. However, with increasing $B$, the electron density loses its influence, resulting in almost vertical lines of equal neutrino rate at $B=\SI{25}{G}$. Here the proton synchrotron emission is supplying the system with a sufficient amount of seed photons for the photohadronic interactions. However, electron synchrotron emission might still be needed to model the low energy emission of a given source.

An often cited quantity is the number of neutrinos emitted per high-energy photon. Either this value is set to a fixed rate, assuming all Fermi photons originating from $\pi_0$-decay, or power-law spectra with a fixed ratio are integrated over the Fermi and IceCube energy range, respectively.
Due to the super position of various processes (see Fig.~\ref{fig:base_sed}) such approaches might lead to large errors with non-linear dependencies on the source parameters.

\begin{figure}[ht]
  \centering
  \includegraphics[width=\textwidth]{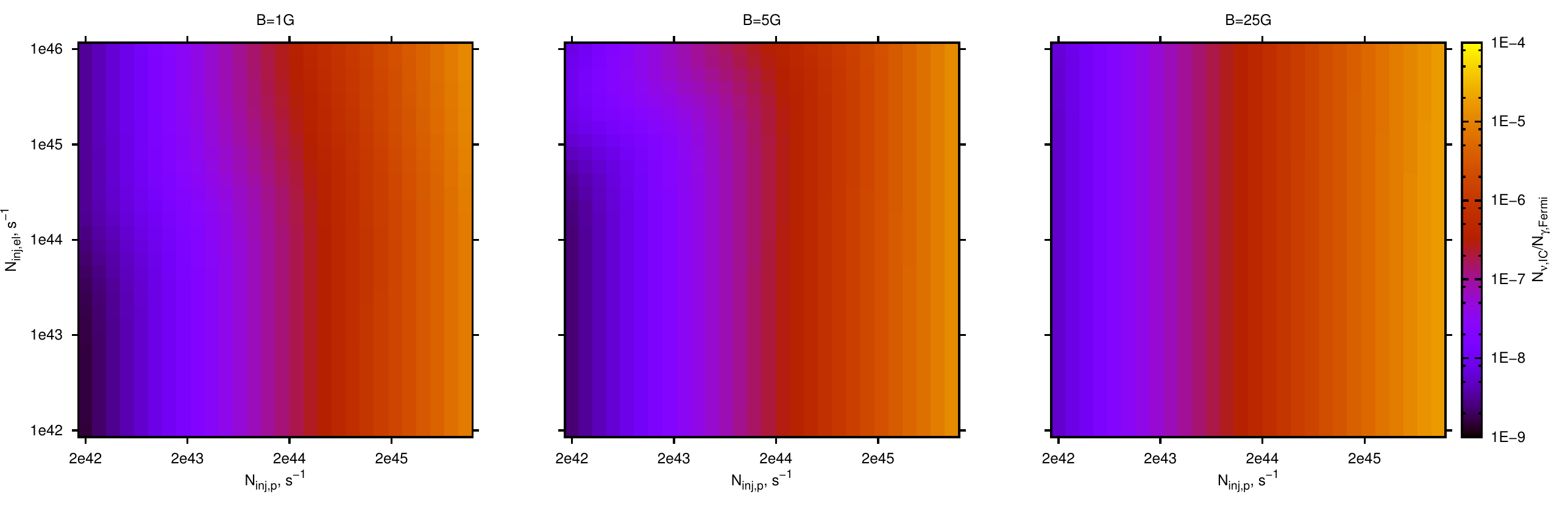}
  \caption{Neutrino count in the energy range of IceCube (both muon and electron neutrinos) over the photon count in the Fermi energy range.}
  \label{fig:icecube_nu_count_over_fermi_photon_count}
\end{figure}

Figure~\ref{fig:icecube_nu_count_over_fermi_photon_count} shows this ratio for our model source.
It is important to note that, in contrast to the $K$ factor, values stay several orders of magnitude below unity which, in the case of \textit{3C 279}, can be attributed to the proton synchrotron emission dominating the Fermi range. Nevertheless this parameter stays a good proxy for the hadronness of a source, although its value will strongly depend on the dominance and shape of the proton synchrotron emission.

\begin{figure}[ht]
  \centering
  \includegraphics[width=\textwidth]{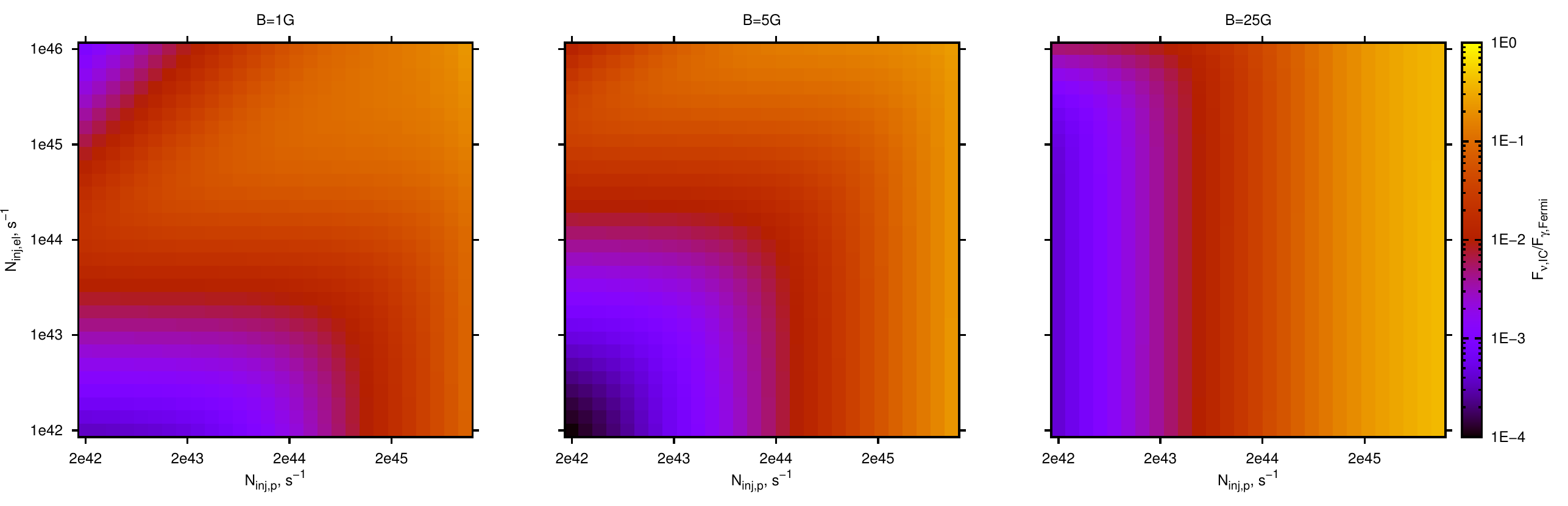}
  \caption{Neutrino energy flux in the energy range of IceCube (both muon and electron neutrinos) over the photon energy flux in the Fermi energy range.}
  \label{fig:icecube_nu_flux_over_fermi_photon_flux}
\end{figure}

When plotting the ratio of energy fluxes instead of current densities (Fig.~\ref{fig:icecube_nu_flux_over_fermi_photon_flux}) the influence of the magnetic field is much more pronounced.
For low $B$ and proton density as well as high electron density the source becomes dominated by inverse Compton scattering, leading to a higher photon flux in the Fermi range without altering the production of neutrinos (compare upper left corner of the left panel of Fig.~\ref{fig:icecube_nu_flux_over_fermi_photon_flux}). With an increasing magnetic field the flux ratio plot approaches the form of the previous plot. The increase of the color scales is due to the energy range of IceCube being at much higher values than the Fermi range.

\begin{figure}[ht]
  \centering
  \includegraphics[width=\textwidth]{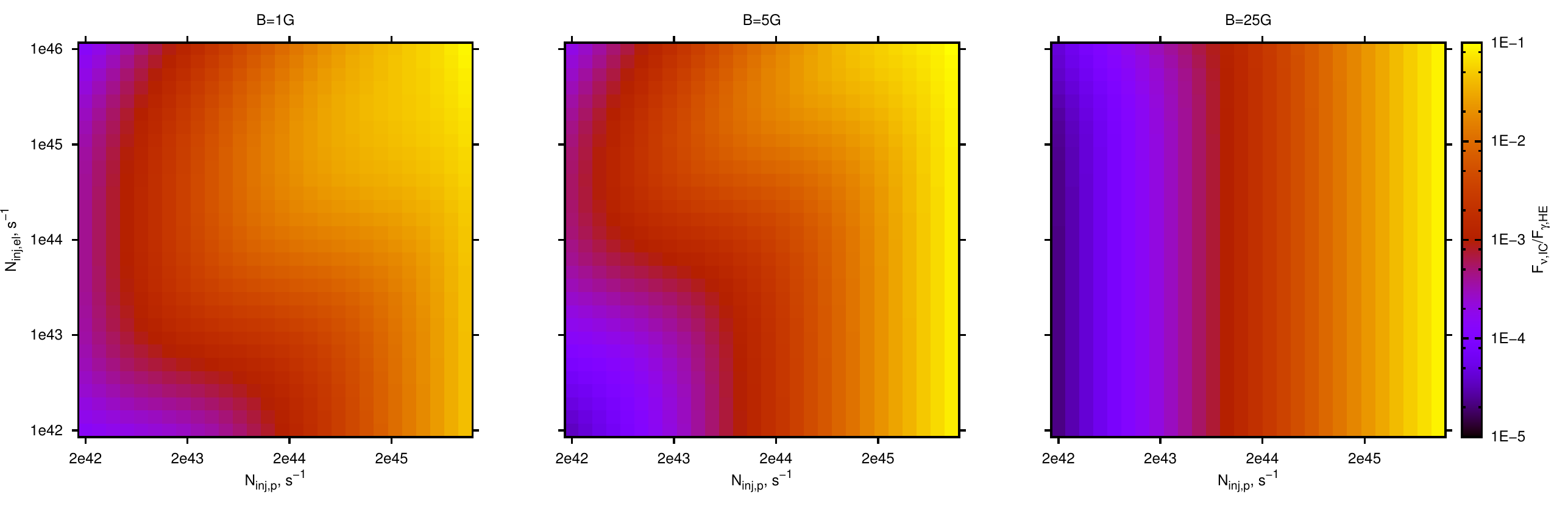}
  \caption{Neutrino energy flux in the energy range of IceCube (both muon and electron neutrinos) over the photon energy flux between the minimum of the SED (between the two peaks) and the highest energies.}
  \label{fig:icecube_nu_flux_over_he_photon_flux}
\end{figure}

Of importance more from a theorists rather than an observers point of view is the ratio of IceCube neutrinos over all high-energy photons, shown in Fig.~\ref{fig:icecube_nu_flux_over_he_photon_flux}.
These plots largely follow Fig.~\ref{fig:icecube_nu_flux_over_fermi_photon_flux}, especially for high magnetic fields. This underlines the assessment of the Fermi range being a good proxy for the neutrino production rate, in otherwise similar sources. If the source does not hold a large proton density and only moderate number of primary electrons, this plot will show a much more structured picture due to the complex interaction of processes in the high-energy regime.

\begin{figure}[ht]
  \centering
  \includegraphics[width=\textwidth]{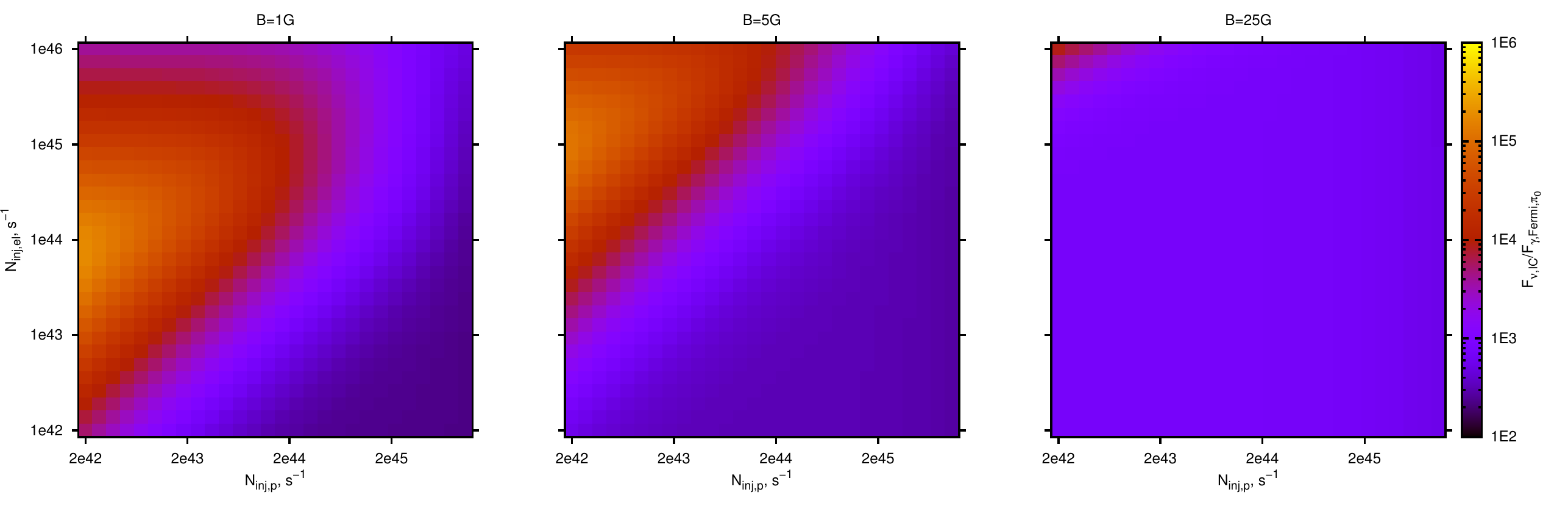}
  \caption{Neutrino energy flux in the energy range of IceCube (both muon and electron neutrinos) over the photon energy flux originating from the decay of $\pi_0$s in the Fermi energy range.}
  \label{fig:icecube_nu_flux_over_fermi_pizero_flux}
\end{figure}

\begin{figure}[ht]
  \centering
  \includegraphics[width=\textwidth]{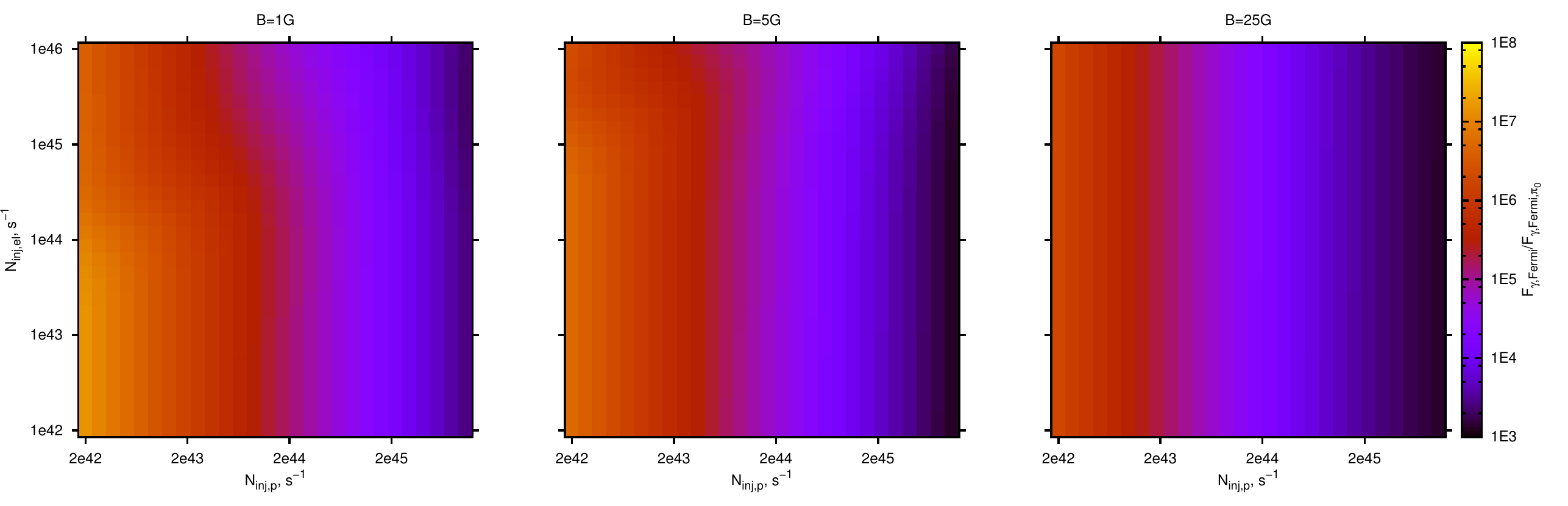}
  \caption{Ratio of the total photon energy flux and the energy flux originating from $\pi_0$ decay integrated over the Fermi energy range.}
  \label{fig:fermi_photon_flux_over_fermi_pizero_flux}
\end{figure}

Finally, Figs.~\ref{fig:icecube_nu_flux_over_fermi_pizero_flux},\ref{fig:fermi_photon_flux_over_fermi_pizero_flux} emphasize the statement, that the assumption of the $\pi_0$s decay playing an important roll in Blazars does not hold.

Naturally the production rate of $\pi_0$-decay-photons is  strongly correlated to the neutrino rate, which however is irrelevant as the former is no observable.
On the contrary, if one would attribute the Fermi-flux to the  $\pi_0$-decay alone, the resulting error in the estimated neutrino flux would be at least four order of magnitude.
This can be seen from the comparison of the color-bar scales of Figs.~\ref{fig:icecube_nu_flux_over_fermi_photon_flux} and~\ref{fig:icecube_nu_flux_over_fermi_pizero_flux}.

This is largely because of the small real conribution of the $\pi_0$ photons to the Fermi flux.
Even for the largest densities the contributions do not go beyond $10^{-3}$ (Fig.~\ref{fig:fermi_photon_flux_over_fermi_pizero_flux}) and show very little dependence on the magnetic field.

\section{Discussion}
\label{sec:discussion}
\subsection{Model restrictions}
As mentioned in section~\ref{sec:model} the chosen two-zone geometry is equivalent to  a one-zone geometry in the steady-state case. On the other hand, a more advanced spatially-resolved model will only affect energies well below the high-energy regime, most notably the radio-regime~\citep{2016ApJ...829...56R}.

The most severe restriction of our model is the absence of proton cooling caused by photo-hadronic interactions. In a large parameter regime, where losses are dominated by synchrotron radiation, this will have no affect on the validity of our results. However, if, for extremely high injection rates, the photo-hadronic interactions produce a sufficient amount of photons to self-propel the process, a closed loop will form. In this case, energy conservation is no longer obeyed and the model would become unstable.
\subsection{Interpretation of results}
The analysis of the performed parameter scan shows a diverse, but still very clear picture.

The similarity of Figs.~\ref{fig:icecube_nu_flux_over_fermi_photon_flux} and~\ref{fig:icecube_nu_flux_over_he_photon_flux} both in structure and overall scale show that the Fermi-flux is still a representative observable for the high energy emissions of a particular source.
However, the mitigation between photons and neutrinos is driven very differently than often assumed:
Although many simplified assumptions on the photohadronic chain hold even for multi-pion production, the simple fact of other photon emission processes having a significant contribution to the Fermi flux will result in large deviations from predictions based on those models.

As can be seen from the SED in Fig.~\ref{fig:base_sed}, the $\pi_0$ photons are usually produced at much higher energies, leaving the Fermi range dominated by either proton synchrotron or pair cascade emission, rendering the photon flux in that energy range highly parameter dependent.
This crucial difference will lead to a more complex dependency on parameters, mainly those influencing the proton synchrotron emission.
The resulting errors for the neutrino-flux estimates can easily reach several orders of magnitude.

However, at least for large magnetic fields, once  there is an estimate for the high-energy proton content of the source, a reliable estimate for the neutrino flux can be computed without a full fit of the SED (compare the vertical symmetry in the right panel of Fig.~\ref{fig:icecube_nu_flux_over_fermi_photon_flux}).

\section{Conclusion}
\label{sec:conclusion}
The work presented shows that there is no bijective mapping of the high-energy photon and neutrino properties for a given source, i.e. independent of any parameter changes. Especially variations of the injected proton density will directly influence the ratio of the emitted neutrino flux to the high-energy photon flux, with even more complex dependencies in lower magnetic fields.

This raises concerns on the reliability of many neutrino flux estimates, especially when involving a large number of sources. Since for such stacking analyses it is not easily possible to extract any source parameters, they might show significant systematic errors that are usually neglected.

In contrast, for the analysis of a small number of sources it would be feasible to get much more precise estimates for the neutrino flux from the shape of the SED. Even a simple modeling of the main contributors would yield parameters like the magnetic field strength and the proton density, which in turn could be used to derive neutrino fluxes using for example the plots presented in this work.
Such an approach would at least require measurements of the spectral indices, fluxes and cut-off frequencies for both peaks of the SED.

Precise models, however, would require flux and slope in five different wave bands to yield constraints for all 10 parameters of the model.
Judging from our experience, reasonable model parameters can be calculated when the electron synchrotron emission can be constrained and when sufficient data for the  HE flux slope is available.

For strong and highly variable sources it might also be worthwhile to employ time-dependent models. Photohadronic processes will result in distinct variability patterns as was shown by \cite{2015A&A...573A...7W} and also allow to investigate the underlying physical processes leading to the acceleration of particles in the first place.

\section*{acknowledgments}
F.S. acknowledges support from NRF through the MWL program. This work is based upon research supported by the National Research Foundation and Department of Science and Technology. Any opinion, findings, and conclusions or recommendations expressed in this material are those of the authors and therefore the NRF and DST do not accept any liability in regard thereto.

\bibliographystyle{elsarticle-num}

\bibliography{apj-jour,ref}

\end{document}